\documentclass[12pt]{article}
\usepackage{amssymb,amsmath}

\newtheorem{theorem}{Theorem}[section]

\newtheorem{proposition}[theorem]{Proposition}

\newenvironment{proof}[1][Proof]{\begin{trivlist}
\item[\hskip \labelsep {\bfseries
#1}]}{\end{trivlist}}

 \newcommand{\qed}{\nobreak
\ifvmode \relax \else
\ifdim\lastskip<1.5em \hskip-\lastskip
\hskip1.5em plus0em minus0.5em \fi \nobreak \vrule height0.75em width0.5em
depth0.25em\fi}
\newcommand{\E}{\textup{\bf E}}
\def\hook{\hbox to 15pt{\vbox{\vskip 6pt\hrule width 6.5pt height 1pt}
         \kern -4.0pt\vrule height 8pt width 1pt\hfil}}

\begin{document}

\begin{centering}

{\huge Functionals and the Quantum Master Equation}

\renewcommand{\thefootnote}{\fnsymbol{footnote}}
\vspace{1.4cm}
{\large Ronald O. Fulp}$^{\dagger}$

\bigskip

\bigskip

SUBMITTED FEB 19 2005 INTERNATIONAL JOURNAL OF THEORETICAL PHYSICS

\vspace{.5cm}
\vspace{.5cm}
\rm

\begin{abstract} The quantum master equation is usually formulated in terms of functionals of the components of mappings (fields in physpeak) from a space-time manifold $M$ into a finite-dimensional vector space. The master equation is the sum of two terms one of which is the anti-bracket (odd Poisson bracket) of functionals and the other is the Laplacian of a functional. Both of these terms seem to depend on the fact that the mappings on which the functionals act are vector-valued. It turns out that neither the Laplacian nor the anti-bracket is well-defined for sections of an arbitrary vector bundle. We show that if the functionals are permitted to have their values in an appropriate graded tensor algebra whose factors  are the dual of the space of smooth functions on $M,$ then both the anti-bracket and the Laplace operator can be invariantly defined.  This permits one to develop the Batalin-Vilkovisky approach to BRST cohomology for functionals of sections of an arbitrary vector bundle.
\end{abstract}

\end{centering}
\vspace{.5cm}

\rm

 \indent{\em   Key Words: quantum master equation, functionals, antibracket, \\
 \indent Laplacian, Batalin-Vilkovisky           }

\rm
\vspace{.5cm}
 \rm

\rm
\vspace{.5cm}
 \rm

$^{\dagger}$ Department of Mathematics, North Carolina State University, \\
 \indent Raleigh, NC 27695-8205,{ \it E-mail: fulp@math.ncsu.edu}\\

\newpage

\section{Introduction} The quantum master equation first appeared as a necessary condition for quantizing certain field theories using the path-integral formalism and BRST cohomological methods \cite{HT92}. Present formulations of the equation use the Batalin Vilkovisky approach to BRST theories making it possible to apply it to a wider class of field theories. 

Among other things the quantum master equation has been responsible for inspiring generalizations of both symplectic geometry and geometric formulations of Laplace's equation to structures on supermanifolds such as odd Poisson structures and odd Laplacians, \cite{Sw93,K91,KV02} and many others. Some of these developments have focussed on finite dimensional theories although the motivating field theories are infinite dimensional.

It is our purpose here to formulate the quantum master equation in a language which allows the fields of a physical theory to be sections of an arbitrary vector bundle $W$. It is our intent to keep the language as close to that used by working physicists as possible. 

This analysis holds no surprises when the bundle is trivial, however in the more general case when $W$ is not trivial, we find that the usual anti-bracket of functionals of sections of the vector bundle $W\rightarrow M$ has no obvious invariant meaning.  We show how to modify the usual idea of a functional by allowing it to be tensor-valued and in this way we obtain an an invariant anti-bracket. This modified functional requires certain extra conditions, a kind of ``equivariance" and a ``reduction" condition on an open cover possessing a partition of unity. With these conditions we can define an anti-bracket and a Laplace operator for these  tensor-valued functionals and consequently a well-defined quantum master equation.

We do not possess a complete understanding of the conditions required for``equivariance" and ``reduction" but we suspect that cohomological obstructions exist in general. This is beyond the scope of the present work which seeks to obtain a correct invariant formulation of the quantum master equation.

\section{Physical and Mathematical preliminaries}

We consider physical theories for which the fields of the theory are sections of a vector bundle $V\rightarrow M$ over a manifold $M$ representing either space or space-time. When $V=M\times V_0$ for some finite-dimensional vector space $V_0$ the fields of the theory may be regarded as mappings $\hat \phi $ from $M$ into $V_0$ and the corresponding sections of $V\rightarrow M$ the mappings  $\phi:M\rightarrow V$ given by $\phi(x)=(x,\hat\phi(x)), x\in M.$ When $M$ is contractible as is the case when $M$ is three-dimensional Euclidean space or Minkowski space, $V$ is necessarily trivial. In general, it is not. A theory such as Yang-Mills theory may be formulated either over Minkowski space or over other manifolds such as spheres or torii depending on boundary conditions. When $M$ is contractible a Yang-Mills field may be identified as a section of the trivial vector bundle $T^*M\otimes g$ where $g$ is the Lie-algebra of the structure group $G$ of the theory.  In general a Yang-Mills field is a connection on a principal fiber bundle $P\rightarrow M$ with structure group $G$. The space of connections is an affine space, but if one fixes a connection $\omega_0,$ then every field (connection) assumes the form $\omega_0+\tau$ where $\tau$ is a tensorial one-form $\tau:TP\rightarrow  g.$ Recall however that if $K=P \times_G g$ is the bundle associated to $P$ and the adjoint action of $G$ on its Lie algebra $g,$ then the space of all  tensorial one-forms $\tau:TP \rightarrow  g$ is in one-to-one correspondence with the space of  mappings $\hat \tau :TM \rightarrow K$  such that, for $x\in M,$ $\hat\tau_x$ maps $T_xM$ into the fiber $K_x$ of $K$ over $x$ (see \cite {KN} pages 75 and 76).  It follows almost immediately that  the space of  tensorial  one-forms  $\tau:TP \rightarrow g$ is in one-to-one correspondence with the space of sections of the vector bundle $T^*M\otimes K \rightarrow M.$ In this way the space of gauge fields may be represented as sections of a vector bundle and consequently falls within the scope of the present paper.  

To implement gauge symmetries of a field theory ghosts, ghosts of ghosts, ect., must be introduced. This gives rise to a new vector bundle $E\rightarrow M$ described briefly below. The notion of an anti-bracket requires new fields called anti-fields which we identify as sections of the dual bundle $E^*\rightarrow M$ (with appropriate parities assigned to its fibers).  Finally, the functionals we consider are mappings from the set of all sections of the bundle $E\oplus E^* \rightarrow M$ into the field ${\bf C}$ of complex numbers.

Since we will be concerned with various bundles $V,E,E^*,E\oplus E^*$ over $M$ we develop our notation and ideas on a generic vector bundle $W\rightarrow M$ specializing $W$ when necessary.

Let
$J^\infty W$ be the infinite jet bundle of $W.$ The restriction of the infinite
jet bundle over  an appropriate open set $U\subset M$ is trivial with fiber an
infinite dimensional vector space $V^\infty$.  The bundle \begin{eqnarray*}
\pi^\infty : J^\infty W_U=U\times V^\infty \rightarrow U \end{eqnarray*} then
has coordinates given by \begin{eqnarray*}
(x^i,u^a,u^a_i,u^a_{i_1i_2},\dots,). \end{eqnarray*} We use multi-index notation
and the summation convention throughout the paper. If $j^{\infty}\phi$ is the
section of $J^{\infty}W$ induced by a section $\phi$ of the bundle $W$, then
$u^a\circ j^{\infty}\phi=u^a\circ \phi$ and $$u^a_I\circ j^{\infty}\phi=
(\partial_{i_1}\partial_{i_2}...\partial_{i_r})(u^a\circ j^{\infty}\phi)$$
where $r$ is the order of the symmetric multi-index
$I=\{i_1,i_2,...,i_r\}$, with the convention that, for $r=0$, there are no
derivatives. For more details see \cite{A96} and \cite{KV98}.

Let $Loc_W$ denote the {\it algebra of local functions} where a {\bf local function} on
$J^\infty W$ is defined to be the pull-back of a smooth function on some finite
jet bundle $J^p W$ via the projection from $J^\infty W$ to $J^p W$. Let
$Loc_W^0$ denote the subalgebra of $Loc_W$ such that $P \in Loc_W^0$ iff
$(j^\infty \phi)^* P$ has compact support for all $\phi \in \Gamma_W$ with
compact support and where $\Gamma_W$ denotes the set of sections of the bundle $W
\to M$. The de Rham complex of differential forms $\Omega^*(J^{\infty}W,d)$ on
$J^{\infty}W$ possesses a differential ideal, the ideal ${ C}$ of contact forms
$\theta$ which satisfy $(j^{\infty}\phi)^* \theta=0$ for all sections $\phi$
with compact support. This ideal is generated by the contact one-forms which,
in local coordinates, assume the form $\theta^a_J=du^a_J-u^a_{iJ}dx^i$.

Now let $C_0$ denote the set of contact one-forms of {\em order zero}. Contact
one-forms of order zero satisfy $(j^{1}\phi)^*(\theta)=0$ and, in local
coordinates, their generators assume the form $\theta^a= du^a-u^a_idx^i$. Let 
$\Omega ^{n,1}$ denote the subspace of $\Omega ^{n+1}(J^{\infty}W)$ which is
locally generated by the forms $\{(\theta ^a \wedge d^nx)\}$ over $Loc_W$.
Notice that both
$C_0$ and $\Omega ^{n,1} = \Omega ^{n,1}(J^ \infty W)$ are modules over
$Loc_W$.  Let $\nu$ denote a fixed 
volume element on $M$  and notice that in local coordinates $\nu$ takes the form
$\nu = f d^nx = f dx^1 \wedge dx^2 \wedge ... \wedge dx^n$ for some function
$f: U \to {\mathbf R}$ where $U$ is a subset of $M$ on which the $x^i$'s are
defined.  Finally, let 
$\Omega ^{n,0}$ denote the subspace of $\Omega ^{n+1}(J^{\infty}W)$ which is
locally generated by the volume $\nu$ over $Loc_W$.

Define the operator $D_i$ (total derivative) by $\displaystyle D_i =
\frac{\partial}{\partial x^i} + u^a_{iJ}\frac{\partial}
{\partial u^a_J}$ (recall that we assume the summation convention, i.e.,
the sum is over all $a$ and multi-index $J$). For $I=\{i_1i_2\cdots i_k\}$ where $k>0$, $D_I$ is defined by $D_I=D_{i_1}\circ D_{i_2} \circ \cdots \circ D_{i_k}$. If $I$ is empty $I=\{\}$ then $D_I$ is just multiplication by 1. We also define $(-D)_I = (-1)^{|I|} D_I$. Recall that the
Euler-Lagrange operator
maps
$\Omega ^{n,0}(J^\infty W)$ into $\Omega ^{n,1}(J^\infty W)$ and is
defined in local coordinates
by $$\E (P \nu)=\E _a(P)(\theta ^a \wedge \nu)$$ where $P \in Loc_W, \nu$
is the
volume
form on the base manifold $M$,
and the
components $\E_a(P)$ are given by $$\E_a(P)=(-D)_I(\frac{\partial P}
{\partial u^a_I}).$$ For simplicity of notation we may use $\E(P)$ for $\E
(P\nu)$. When $W=E\oplus E^*$ for some bundle $E$ we denote the 
Euler operator acting on a local function $P\in Loc_W$ by $\widetilde \E_aP.$ 
The components of this operator are of two types. Certain of them refer
to coordinates on the vector bundle $E$ and the others to coordinates
on its dual bundle $E^*.$ This complicates the notation and will be dealt
with as the need arises. We drop the bold type for the Euler operator 
throughout the remainder of the paper as there will be little confusion
in doing so.

We also will have occasional use for the  the space of horizontal forms  $\Omega ^{n-1,0}$ which is locally generated by the forms $\partial_{\mu} \hook \nu$ over $Loc_W.$ The corresponding horizontal differential $d_H:\Omega ^{n-1,0}\rightarrow \Omega ^{n,0}$ is defined by 
\newline $d_H(j^{\mu}( \partial_{\mu} \hook \nu)) =( D_{\mu}j^{\mu})\nu.$

To say that ${\cal A}$ is a {\bf local functional} means that ${\cal A}$ is a mapping from the space of $\Gamma=\Gamma_W$ of sections of $W\rightarrow M$ into ${\bf C}$ such that for some local function $A:J^{\infty}W\rightarrow {\bf C},$
$${\cal A}(\phi)=\int_M[A\circ j\phi]\nu$$ 
for all $\phi\in \Gamma.$  

{\it Here, as above, $\nu$ is a volume on $M$ which will remain fixed throughout the paper and for simplicity of notation $j\phi$ will denote  the section of $J^{\infty}W\rightarrow M$ induced by $\phi$ (denoted by $j^{\infty}\phi$ above). }

A field Lagrangian is a local function $L$ on $J^{\infty}V$ where $V\rightarrow M$ is the bundle whose sections are the fields of the theory. Its corresponding functional $S:\Gamma_V\rightarrow {\bf C}$ given by 
$$S(\phi)=\int_M[L\circ j\phi]\nu$$
is the {\bf action} of the theory.

We say that $L$ possesses gauge freedom if there is a parameter space $\Lambda$ and a mapping $\delta$ from $\Lambda$ into the space of generalized vector fields on $J^{\infty}V$ subject to certain properties. First, for $\xi\in\Lambda,$ $\delta(\xi)$ is an evolutionary vector field.  Such fields may be written as $$\delta(\xi)=\delta^a(\xi) \frac{\partial }{\partial u^a}$$ where $(x^{\mu},u^a_I)$ are local coordinates on $J^{\infty}V$ adapted to a chart $(x^{\mu})$ on $M$ and where $\delta^a(\xi)\in Loc_V.$ Moreover it is required that such a vector field have a unique prolongation $Pr(\delta(\xi))$ to $J^{\infty}V$ and that this prolongation satisfy the condition: $Pr(\delta(\xi))(L)=d_H\alpha$ where $d_H$ is the horizontal differential on the variational complex on $J^{\infty}V$  \cite{O86,BFLS98, AF05}. In local coordinates this means that $Pr(\delta(\xi))(L)$ is the product of a divergence with the volume $\nu.$ This is equivalent to saying that the action is preserved under the flow of $Pr(\delta(\xi))$ on $J^{\infty}V.$ Generally, $\Lambda$ may be identified as a linear subspace of the space of all sections of a vector bundle $B\rightarrow J^{\infty}V$ (see \cite{GMS05}). This bundle is dependent on the Lagrangian of the theory. In general, roughly speaking, gauge parameters are functions of points of M, of the fields of the theory, and of the derivatives of the fields. In simpler cases they are functions only of points of $M$ and in this case $\Lambda$ is a subspace of the space of sections of a vector bundle $B\rightarrow M.$ For simplicity we restrict our attention to this case.

In this case one introduces a new vector bundle $C^1\rightarrow M$ whose fibers are the same as those of $B\rightarrow M$ but with reversed parity. Its sections are called ghost fields. If the gauge symmetries are reducible (see \cite{HT92}  for relevant definitions) there should, in principle,  exist additional bundles 
$$C^2\rightarrow M,  C^2\rightarrow M, \dots, C^k\rightarrow M$$
depending on the order of reducibility. The author has not seen this explicitly done  for the general case and in fact certain rank conditions are undoubtedly required \cite{BS03}. Sections of $C^2\rightarrow M$ are called ghosts of ghosts and similar language is used for higher orders of reducibility. 

We define the vector bundle $E$ by 
$$E=V\oplus C^1\oplus C^2\oplus \cdots \oplus C^k \longrightarrow M.$$
Sections of $E\rightarrow M$ include all fields and ghost fields. For details regarding parity see Henneaux and Teitelboim \cite{HT92}.

We should point out that an alternative approach to incorporating ghosts, ghosts of ghosts,  ect., makes use of a sheaf-theoretic approach \cite{GMS05} which in many ways is preferable, but we want to keep the language closer to that of \cite{HT92} and of many practitioners of the theory.

{\it From this point on we refer to sections of $E$ as being fields. We introduce the bundle $E^*\rightarrow M$ dual to $E$ but assign to each fiber $E^*_x$ a parity which is opposite that of $E_x,  x\in M.$ Sections of this new bundle will be called {\bf anti-fields.}}

Sections of the bundle $E\oplus E^* \rightarrow M$ will be denoted by 
$\tilde \phi=(\phi,\phi^*)$ where $\phi$ is a section of $E\rightarrow M$ and $\phi^*$ is a section of $E^*\rightarrow M.$ If $(x^{\mu},u^a)$ and $(x^{\mu},u^*_b)$ are charts of $E$ and $E^*,$ respectively, which are adapted to the chart $(x^{\mu})$ of $M$ then $(x^{\mu},u^a,u^*_b)$ is a chart of $E\oplus E^*$ adapted to $(x^{\mu}).$ If the coordinate $u^a$ has parity $\varepsilon^a,$ then the coordinate $u^*_a$ has parity $\varepsilon^*_a=\varepsilon^a+1.$

\section{ Anti-brackets and the Laplacian}

Roughly speaking, a functional  is a mapping ${\cal A}$ from the space $\Gamma=\Gamma_W$ of all sections of a vector bundle $W\rightarrow M$ into the field ${\bf C}$ of complex numbers. We will assume that ${\cal A}$ is smooth in the sense we now describe. Any definition of derivative of a function defined on a vector space such as $\Gamma$ seems to require a topology on that space such that the operations are continuous. One can obtain a meaningful concept of derivative on a Fre$\acute  c$het space. There are many ways to obtain such a topology on the space $\Gamma$ of sections of a vector bundle. We require no particular topology on $\Gamma$ except that it provide a reasonable class of functions which are differentiable. We require the following properties of the derivative and any definition satisfying these properties will be adequate for our purposes here. Let $S$ be any  topological vector space over ${\bf C}.$ If ${\cal A}$ is a mapping from $\Gamma$ into $S$ and $\phi\in \Gamma,$ then the derivative, $D_{\phi}{\cal A}$ of ${\cal A}$ at $\phi,$ is required to be a linear mapping from $T_{\phi}\Gamma=\Gamma$ into $S$ subject to the following properties:

(1) If $\lambda \rightarrow \phi(\lambda)$ is a smooth curve from an interval into $\Gamma,$  then $$\frac {d}{d \lambda}[\phi(\lambda)](x)=\frac {d}{d \lambda}[\phi(\lambda)(x)], \quad x\in M.$$

(2) If $\lambda \rightarrow \phi(\lambda)$ is a smooth curve from an interval into $\Gamma,$  then 
$$(D_{\phi(\lambda)}{\cal A})(\phi(\lambda))=\frac{d}{d\lambda}{\cal A}(\phi(\lambda)).$$

(3) If $C^{\infty}M$ acts continuously and linearly on $S$ and ${\cal A}(f\phi)=f{\cal A}(\phi)$ for $f\in C^{\infty}M, \phi\in \Gamma,$ then
$$\frac{d}{d\lambda}{\cal A}(\phi_0+\lambda f\delta)= fD_{\phi_0}{\cal A}(\delta)$$
for $f\in C^{\infty}M, \phi_0, \delta \in \Gamma, \lambda \in{\bf R}.$

\bigskip

A {\bf functional}  is a smooth (infinitely differentiable) mapping ${\cal A}$ from $\Gamma$ into ${\bf C}.$ 
\bigskip

\noindent{\bf Remark:} Recall that if $\lambda \rightarrow \phi(\lambda)$ is a (smooth) curve in $\Gamma,$ such that $\phi(0)=\phi_0, \quad \frac {d}{d\lambda}\phi(\lambda)|_{\lambda=0}=\delta \in T_{\phi_0}\Gamma ,$ then 
$$D_{\phi_0}{\cal A}(\delta)=\frac {d}{d\lambda}\int_M[A\circ j\phi] \nu =\int_M[E(A)\circ j\phi](\delta)\nu$$
where $E$ is the usual Euler operator regarded as a mapping from $\Omega^{n,0}$ into $\Omega^{n,1}$ and where we make the usual assumptions regarding the vanishing of surface terms at infinity. Here $(E(A)\circ j\phi)(\delta) = (E_a(A)\circ j\phi)\delta^a.$ Notice that $\delta$ is tangent to $\Gamma$ at $\phi_0$ but it may also be identified as a vector field along $\phi_0,$ i.e.,
$\delta(x)\in T_{\phi(x)}J^{\infty}W$ for each $x\in M.$

Let ${\cal F}={\cal F}_{\Gamma}$ denote the linear space of all functionals on $\Gamma=\Gamma_W.$ Observe that it is an {\it algebra} relative to the usual pointwise operations: 
$$({\cal A}_1+{\cal A}_2)(\phi)={\cal A}_1(\phi)+{\cal A}_2(\phi), (c{\cal A})(\phi)=c{\cal A}(\phi), ({\cal A}_1{\cal A}_2)(\phi)={\cal A}_1(\phi){\cal A}_2(\phi),$$  
for ${\cal A},{\cal A}_1,{\cal A}_2\in{\cal F}, c\in {\bf C},\phi\in \Gamma.$
Also observe that the space of local functionals is a subspace of ${\cal F}$ but is {\it not} a subalgebra. Thus the space of local functionals is not a Poisson algebra relative to the anti-bracket.

The quantum master equation is 
$$\Delta{\cal A}+\frac{i}{\hbar}({\cal A},{\cal S})=0$$
where ${\cal A}$ is a functional and ${\cal S}$ is a {\it local} functional which is an appropriate homological deformation of the action functional ${\cal S}_0$ defined by the Lagrangian $L$ of the field theory:  ${\cal S}_0(\phi)=\int_M[L\circ j\phi] \nu, \quad \phi\in \Gamma$ (see \cite{HT92}
for details).
Here $(\cdot,\cdot)$ is the anti-bracket of functionals and $\Delta$ is the Laplace operator. It is precisely these concepts which are the primary focus of this paper. One has a ``reasonable" definition of the anti-bracket and the Laplace operator for field theories for which the fields have their values in a finite-dimensional vector space, but not for fields which are vector-bundle valued. It is our intent to rectify this situation.

Recall that if the fields have their values in the vector space $V_0,$ then the anti-bracket is usually defined for functionals ${\cal A},{\cal B},$ by:
$$({\cal A},{\cal B})= 
\frac {\delta^R{\cal A}}{\delta \phi^a}\frac {\delta^L{\cal B}}{\delta \phi^*_a}-
\frac {\delta^R{\cal A}}{\delta \phi^*_a}\frac {\delta^L{\cal B}}{\delta \phi^a}$$ 
which involve left and right partial functional derivatives (see \cite{HT92}, page 417). Similarly, the Laplacian is also defined in terms of derivatives of this kind: 
$$\Delta=
(-1)^{\varepsilon_a+1}\frac {\delta^R}{\delta \phi^a}\frac {\delta^R}{\delta \phi^*_a}.$$
The $\phi^a, \phi^*_a$ which appear in these formulas refer to the components of the fields and anti-fields relative to the choice of a basis $\{e_a\}$ in $V_0$ and its corresponding dual basis $\{e^b\}$ of $V_0^*,$ the dual space of $V_0$ (anti-fields are identified as mappings from the space-time manifold $M$ into $V_0^*$ with appropriate parity assignments). If one chooses another basis $\{\bar e_a\}$ along with its dual basis $\{\bar e^b\},$ then the definitions given above are basis independent since 
$e_b=A_b^ae_a $ for some matrix $A$ of {\it scalars} and $e^b=B^b_ae^a$ where the matrix $B$ is the inverse of $A.$

More generally, by analogy with ordinary partial derviatives, we define $\frac{\delta^L}{\delta \psi^a}$ by
$$\frac{\delta^L\widetilde {\cal A}}{\delta \psi^a}(\phi_0)=\frac{d}{d\lambda}\widetilde {\cal A}(\phi_0+\lambda e_a)|_{\lambda=0}=D_{\psi_0}\widetilde {\cal A}(e_a)$$
where $\{e_a\}$ is a basis of local sections of $W\rightarrow M$ defined on some open subset $U$ of $M.$ Here $\psi\in \Gamma$ has the form $\psi=\psi^ae_a, \psi^a\in C^{\infty}U.$  This partial functional derivative is only defined on local sections of $W\rightarrow M$ which brings up issues dealt with in detail later in the paper. Right functional derivatives are defined by the following relation:
$$\frac{\delta^L\widetilde {\cal A}}{\delta \psi^a} =-(-1)^{\varepsilon(\widetilde{\cal A})\varepsilon(\psi^a)} \frac{\delta^R\widetilde {\cal A}}{\delta \psi^a}.$$

In case $W=E\oplus E^*,$ sections $\phi$ of $E|U\rightarrow U$ and $\phi^*$ of $E^*|U\rightarrow U$ can be written as $\phi=\phi^ae_a,\phi^*=\phi^*_be^b$ as in the case of vector-valued fields described above. In this case, where the fields are bundle-valued, if  $\{\bar e_a\}$ and $ \{\bar e^b\} $ is another choice of local bases of the restrictions of $E$ and $E^*$ to $U,$ respectively, then they are related to each other as above: 
$\bar e_a=A_a^be_b, \bar e^b=B^b_ae^a$ with one crucial difference, namely that  the entries of the matrices $A$ and $B$ are in $C^{\infty}U.$ 

Partial functional derivatives of a functional of mappings $\phi:M\rightarrow V_0$  satisfy the transformation law $\frac{\delta{\cal A}}{\delta \bar\phi^a}=A_a^b \frac{\delta{\cal A}}{\delta\phi^b}$ for {\it scalars} $A_a^b.$ This presents no difficulty since each term in the transformation law is a complex number (when evaluated at a field). But in the case that the $A_a^b$ are {\it functions} one side of the equation is a complex number while the other side is a function!  It is this issue which we deal with below.

To handle this problem we will modify the concept of functional by permitting it to be ``tensor-valued'' in the sense described below. We also require that these new functionals satisfy a kind of ``equivariance" condition and a ``reduction" condition. These new functionals are useful to encode higher order functional derivatives and so refer to linearizations of an ordinary functional near some field $\phi.$

In that which follows we consider two copies of the dual space $(C^{\infty}M)^*\otimes {\bf C}$ which we denote by $C_0^{\infty}M^*$ and $C_1^{\infty}M^*.$  We say that $\alpha \in C_0^{\infty}M^*$  is {\bf even} and write $\varepsilon (\alpha)=0$ and that $\beta\in C_1^{\infty}M^*$ is {\bf odd} with $\varepsilon(\beta)=1.$ For
$\varepsilon=\varepsilon(k)=(\varepsilon_1,\varepsilon_2,\cdots,\varepsilon_k) $ with $\varepsilon_i\in \{0,1\},$ for each $1\leq i\leq k,$ $C_{\varepsilon}^{\infty}M^*$ will denote the tensor product
$$C_{\varepsilon_1}^{\infty}M^*\hat\otimes C_{\varepsilon_2}^{\infty}M^*\hat \otimes \cdots \hat \otimes C_{\varepsilon_k}^{\infty}M^*.$$
The tensor product is presumed to be graded commutative in the sense that  for $\alpha\in C_{\varepsilon}^{\infty}M^*, \alpha' \in C_{\varepsilon'}^{\infty}M^*,
\varepsilon=(\varepsilon_1,\varepsilon_2,\cdots,\varepsilon_k) , \varepsilon'=(\varepsilon'_1,\varepsilon'_2,\cdots,\varepsilon'_l), $
$$\alpha\hat\otimes \alpha'=
(-1)^{|\varepsilon||\varepsilon'|}(\alpha'\hat\otimes \alpha)\in C_{(\varepsilon,\varepsilon')}^{\infty}M$$
where $|\varepsilon|=(\varepsilon_1+\varepsilon_2+\cdots+\varepsilon_k) (mod  \quad 2),
|\varepsilon'|=(\varepsilon'_1+\varepsilon'_2+\cdots+\varepsilon'_l) (mod \quad 2),$ and
$(\varepsilon,\varepsilon')=
(\varepsilon_1,\varepsilon_2,\cdots,\varepsilon_k,\varepsilon'_1,\varepsilon'_2,\cdots,
\varepsilon'_l).$ The tensor product is also supposed to be linear over $C^{\infty}M$ relative to the module structures we now define. Let $C^{\infty}M$ act on the dual spaces below as follows: define maps
$C^{\infty}M \times C^{\infty}_0M^* \rightarrow C^{\infty}_0M^*,
C^{\infty}M\times C^{\infty}_1M^*\rightarrow C^{\infty}_1M^*$
by $(f\alpha)(g)=\alpha(fg)$ for $f,g \in C^{\infty}M$ and $\alpha$ in either $C^{\infty}_0M^*$ or $C^{\infty}_1M^*.$ Then $C^{\infty}M$ acts linearly on both copies of $C^{\infty}M^*.$ Now extend the action of $C^{\infty}M$ so that it acts on $C_{\varepsilon}^{\infty}M^*,  \varepsilon\in\{0,1\}^k,$ by
$$g(\alpha_1 \hat\otimes \alpha_2  \hat\otimes \cdots \hat\otimes \alpha_k)=
\alpha_1\hat\otimes  \alpha_2 \hat\otimes\cdots \hat\otimes (g \alpha_i)
\hat\otimes \cdots \hat\otimes \alpha_k)$$
for $g\in C^{\infty}M, \alpha_i \in C_{\varepsilon_i}^{\infty}M^*, 1\leq i \leq k.$

Notice that $\hat\otimes$ can be defined in terms of the ordinary tensor product of copies of $C^{\infty}M^*$ via
$$\alpha_1\hat\otimes \alpha_2\hat\otimes \cdots \hat\otimes \alpha_k=
\sum_{\sigma}(-1)^{\kappa(\sigma)}(\alpha_{\sigma (1)} \otimes \alpha_{\sigma(2)}\otimes \cdots\otimes \alpha_{\sigma (k)})$$
where ${\kappa(\sigma)}$ is the Koszul sign of the permutation ${\sigma}.$ Thus if $g=(g_1,g_2,\cdots,g_k)\in C^{\infty}M^k,$ then
$$(\alpha_1\hat\otimes \alpha_2\hat\otimes \cdots \hat\otimes \alpha_k)(g)=
\sum_{\sigma}(-1)^{\kappa(\sigma)}\alpha_{\sigma(1)}(g_1) \alpha_{\sigma (2)}(g_2) \cdots \alpha_{\sigma (k)}(g_k).$$
\bigskip

A {\bf tensor-valued functional} is a smooth mapping $\widetilde {\cal A}$ from $\Gamma$ into $C^{\infty}_{\varepsilon}M^*$ (for some $\varepsilon=\varepsilon(k) \in \{0,1\}^k$) such that 
$\widetilde {\cal A}(g\phi)=g\widetilde {\cal A}(\phi), g\in C^{\infty}M, \phi\in \Gamma.$ In this case we say that $\widetilde {\cal A}$ has parity $\varepsilon(\widetilde {\cal A})=|\varepsilon|$ with an obvious abuse of notation.

Notice that if ${\cal A}:\Gamma\rightarrow {\bf C}$ is a {\it linear} functional, then $\widetilde {\cal A}:\Gamma\rightarrow C^{\infty}_{\varepsilon}M^*$ defined by 
$\widetilde {\cal A}(\phi)(f)={\cal A}(f\phi), f\in C^{\infty}M,\phi\in \Gamma,$ is a tensor-valued functional for $\varepsilon=0$ or $\varepsilon=1.$

It follows that if ${\cal A}:\Gamma \rightarrow {\bf C}$ is {\it any} functional, then, for each $\phi\in \Gamma,$ one has a tensor-valued functional $\widetilde {\cal A}_{\phi}$ defined by $$\widetilde {\cal A}_{\phi}(\delta)(f)=D_{\phi}{\cal A}(f\delta),$$
for $f\in C^{\infty}M, \delta\in \Gamma.$

{\it We refer to the tensor-valued functional $\widetilde {\cal A}_{\phi}$ as the 
{\bf  linearization} of ${\cal A}$ at $\phi\in \Gamma.$}

Notice that if ${\cal A}:\Gamma\rightarrow {\bf C}$ is a local functional, ${\cal A}(\phi)=\int_M[A\circ j\phi] \nu,\phi\in \Gamma, A\in Loc_W,$ then the linearization of ${\cal A}$ near $\phi\in \Gamma$ is given, in local coordinates, by
$$\widetilde {\cal A}_{\phi}(\delta)(f)=\int_M [E_a(A)\circ j\phi](f\delta^a) \nu,$$
$f\in C^{\infty}M, \delta\in \Gamma.$

Recall that the partial functional derivatives were defined  by
$$\frac{\delta^L\widetilde {\cal A}}{\delta \phi^a}(\phi_0)=\frac{d}{d\lambda}\widetilde {\cal A}(\phi_0+\lambda e_a)|_{\lambda=0}=D_{\phi_0}\widetilde {\cal A}(e_a)$$
where $\{e_a\}$ is a basis of local sections of $W\rightarrow M$ defined on some open subset $U$ of $M.$ Here $\phi\in \Gamma$ has the form $\phi=\phi^ae_a, \phi^a\in C^{\infty}U.$ This definition does not work however as $\phi_0+\lambda e_a$  is defined only on $U$ and thus is not in the domain of $\widetilde {\cal A}$ which is $\Gamma_W.$ To get past this difficulty we restrict the class of tensor-valued functionals below, but first we need further notation.

\bigskip

Assume that $\widetilde {\cal A}:\Gamma\rightarrow C^{\infty}_{\varepsilon}M^*, \varepsilon=\varepsilon(k)\in \{0,1\}^k,$ is an arbitrary tensor-valued functional. If ${\cal U}=\{U_{\alpha}\}$ is an open cover of $M,$ we say that $\widetilde {\cal A}$ {\bf reduces} to ${\cal U}$ if and only if there exists a partition of unity $\{f_{\alpha}\}$ subordinate to ${\cal U}$ and a family $\{\widetilde {\cal A}_{\alpha}\}$ of tensor-valued functionals
$$\widetilde {\cal A}_{\alpha}:\Gamma_{\alpha}\rightarrow C^{\infty}_{\varepsilon(k)}U_{\alpha}^*$$
defined on the space $\Gamma_{\alpha}$ of sections of $W|U_{\alpha}\rightarrow U_{\alpha},$ for each $\alpha,$ such that for arbitrary $\phi\in \Gamma $ and 
$g\in C^{\infty}M^k,$

$$\widetilde {\cal A}(\phi)(g)=\sum_{\alpha}\widetilde {\cal A}_{\alpha}(f_{\alpha} \phi)(g|U_{\alpha})$$
where $g|U_{\alpha}=(g_1|U_{\alpha}, g_2|U_{\alpha},\cdots,g_k|U_{\alpha}).$
 
 \bigskip
 {\it Our basic assumption is that there exists a significant class of reducible tensor-valued functionals. It is not clear how restrictive this assumption is. We conjecture that obstructions to existence could be classified cohomologically  but we have not undertaken such a classification.}
 
 \begin{proposition} The linearization $\widetilde {\cal A}_{\phi}$ of an arbitrary {\it local} functional at $\phi \in \Gamma$ reduces to any open cover ${\cal U}$ such that (1) ${\cal U}$  possesses a subordinate partition of unity and (2) $W|U\rightarrow U$ is trivial for each $U\in {\cal U}.$
 \end{proposition}

\begin{proof} Let ${\cal A}$ be any local functional and $\widetilde{\cal A}_{\phi}$ its linearization at $\phi \in \Gamma.$  For $f\in C^{\infty}M, \delta\in \Gamma,$
$$\widetilde{\cal A}_{\phi}(\delta)(f)=\int_M[E(A)\circ j\phi] (f\delta) \nu.$$
Let ${\cal U}=\{U_{\alpha}\}$ be an open cover of $M$ with subordinate partition of unity $\{f_{\alpha}\}.$ For each $\alpha,$ let $A_{\alpha}= A| (\pi^{\infty}_M)^{-1}(U_{\alpha})$ and 
$$\widetilde {\cal A}_{\alpha}(\psi)(g)=
\int_{U_{\alpha}}[E(A_{\alpha})\circ ( j\phi)|(U_{\alpha})](g\psi) \nu$$
for $g\in C^{\infty}U_{\alpha}, \psi\in \Gamma_{\alpha}$ where $\Gamma_{\alpha}$ is the space of sections of $W|U_{\alpha}\rightarrow U_{\alpha}.$
Now observe that 
\begin{eqnarray*} \widetilde{\cal A}_{\phi}(\delta)(g) &=&
\widetilde{\cal A}_{\phi}(\sum_{\alpha}f_{\alpha}\delta)(g)\\  &=&
\int_M\{[E(A)\circ j\phi] (\sum_{\alpha}(gf_{\alpha})  \delta \}) \nu \\ &=&
\sum_{\alpha}\int_{U_{\alpha}}\{[E(A_{\alpha})\circ ( j\phi)|U_{\alpha}]((gf_{\alpha})\delta\})             
 \nu\\ &=&
\sum_{\alpha}\widetilde {\cal A}_{\alpha}(f_{\alpha}\delta)(g|U_{\alpha}).
\end{eqnarray*} 
This concludes the proof of the proposition.
\end{proof}

\bigskip

{\it Assume that ${\cal U}=\{U_{\alpha}\}$ is a fixed open cover of $M$ such that $W|U_{\alpha}\rightarrow U_{\alpha}$ is trivial for each $\alpha$ and such that $\{f_{\alpha}\}$ is a partition of unity subordinate to ${\cal U}.$ For each such cover let ${\cal R}_{\varepsilon}={\cal R}_{\cal U}(\varepsilon)$ denote the  set of all tensor-valued functionals $\widetilde {\cal A}:\Gamma \rightarrow C^{\infty}_{\varepsilon}M^*,  \varepsilon \in \{0,1\}^k,$  which reduce to ${\cal U}.$  Moreover, let ${\cal R}={\cal R}_{\cal U}$ denote the linear space spanned by elements of ${\cal R}_{\varepsilon}$ for various $\varepsilon.$}

\begin{theorem} The set ${\cal R}$ of all tensor-valued functionals which reduce to ${\cal U}$ is a linear space under pointwise operations. It is in fact an algebra with the product  $\hat\otimes$ defined by $(\widetilde{\cal A}\hat\otimes \widetilde{\cal B})(\phi)= \widetilde{\cal A}(\phi) \hat\otimes \widetilde{\cal B}(\phi),$ for $\widetilde{\cal A},\widetilde{\cal B}\in {\cal R} ,$ and $\phi\in \Gamma.$ Moreover, when $W=E\oplus E^*$ there is a well-defined Laplace operator and a well-defined anti-bracket on ${\cal R}$ which is graded skew-symmetric and which satisfies the graded Jacobi identity. 
\end{theorem}

\begin{proof} The proof that ${\cal R}$ is an algebra is straightforward and is left to the reader. To define the anti-bracket requires a number of steps.

First we must deal with the issue of parity. If $\widetilde {\cal A}$ maps $\Gamma$ into $C^{\infty}_{\varepsilon}M^*,$ then we define the parity of $\widetilde {\cal A}$ to be $|\varepsilon|.$ With this definition the product on ${\cal R}$ is obviously graded commutative since that property holds on the range of the mapping $\widetilde {\cal A}.$ Since left and right partial functional derivatives are related to each other by the formula:
$$\frac{\delta^L\widetilde {\cal A}}{\delta \psi^a} =-(-1)^{\varepsilon(\widetilde{\cal A})\varepsilon(\psi^a)} \frac{\delta^R\widetilde {\cal A}}{\delta \psi^a},$$
{\it it is sufficient to work out the necessary invariance properties using left derivatives which we denote as one would a derivative in the bosonic case.}

We now show that if $\tilde \phi=(\phi, \phi^*)$ is a section of $E\oplus E^*\rightarrow M$ where $\phi,\phi^*$ are sections of $E\rightarrow M$ and $E^*\rightarrow M,$ respectively, then
for arbitrary $\widetilde {\cal A}:\Gamma_U\rightarrow C^{\infty}_{\varepsilon}U^*,\widetilde {\cal B}:\Gamma_U\rightarrow C^{\infty}_{\varepsilon'}U^*$
$$\frac{\delta\widetilde {\cal A}}{\delta \phi^a} \hat\otimes \frac{\delta\widetilde {\cal B}}{\delta \phi^*_b}$$ is invariant under change of components of the fields. Here $U$ is an arbitrary element of the open cover ${\cal U}.$ To see this choose bases $\{e_a\}, \{\bar e_a\}$ of sections of $E|U\rightarrow U$ along with the corresponding dual bases $\{e^b\}, \{\bar e^b\}$ of $E^*|U\rightarrow U.$ Let $A$ be a matrix whose entries are functions in $C^{\infty}U$ such that $\bar e_a=A_a^be_b$ and let $B$ denote the inverse of $A$ so that $\bar e^b=B^b_ae^a.$ Now (operating with left derivatives) we have
$$\frac{\delta\widetilde {\cal A}}{\delta \bar \phi^b} (\tilde\phi)=D_{\tilde\phi}\widetilde {\cal A}(\bar e_b)
=D_{\tilde\phi}\widetilde {\cal A}(A_b^a e_a)=A_b^aD_{\tilde\phi}\widetilde {\cal A}(e_a)=
A_b^a \frac {\delta \widetilde {\cal A}}{\delta \phi^a} (\tilde\phi).$$
Notice that we have used the fact that $D_{\tilde\phi}\widetilde{\cal A}$ is linear over $C^{\infty}M$ which is a consequence of the third property of derivatives given at the beginning of this section (recall that $\widetilde {\cal A}$ is linear over $C^{\infty}M$). A similar argument shows that
$$\frac{\delta\widetilde {\cal A}}{\delta \bar \phi^*_b} (\tilde\phi)=B_a^b \frac {\delta \widetilde {\cal A}}{\delta \phi^*_a} (\tilde\phi).$$ It follows that 
$$\frac{\delta\widetilde {\cal A}}{\delta \bar \phi^b} (\tilde\phi)\hat\otimes \frac{\delta\widetilde {\cal A}}{\delta \bar \phi^*_b} (\tilde\phi)=A_b^a \frac {\delta \widetilde {\cal A}}{\delta \phi^a} (\tilde\phi)\hat\otimes
B_a^b \frac {\delta \widetilde {\cal A}}{\delta \phi^*_a} (\tilde\phi)$$
and that
$$\frac{\delta\widetilde {\cal A}}{\delta \bar \phi^b} (\tilde\phi)\hat\otimes \frac{\delta\widetilde {\cal A}}{\delta \bar \phi^*_b} (\tilde\phi)=\frac {\delta \widetilde {\cal A}}{\delta \phi^a} (\tilde\phi)\hat\otimes \frac {\delta \widetilde {\cal A}}{\delta \phi^*_a} (\tilde\phi),$$
where we use the fact that the matrix $A$ is the inverse of $B$ and the fact that the tensor product is linear over $C^{\infty}M.$

Consequently, the combination $$\frac {\delta \widetilde {\cal A}}{\delta \phi^a} (\tilde\phi)\hat\otimes \frac {\delta \widetilde {\cal A}}{\delta \phi^*_a} (\tilde\phi)$$
is independent of the basis of local sections used to define it.

Observe that $\frac {\delta \widetilde {\cal A}}{\delta \phi^a} (\tilde\phi)$ depends both on the functional $\widetilde{\cal A}$ and on the local section $e_a$ used to define it. Thus its parity should depend on both of these. Define
$$\varepsilon(\frac {\delta \widetilde {\cal A}}{\delta \phi^a} (\tilde\phi))=
\varepsilon(\widetilde {\cal A} )+\varepsilon(e_a)=\varepsilon(\widetilde {\cal A}) +\varepsilon^a.$$
We require that if $\tilde \phi=\phi^ae_a+\phi^*_be^b,$ then $\varepsilon(e_a)=\varepsilon(\phi^a)=\varepsilon^a$ and $\varepsilon(e^b)=\varepsilon(\phi^*_b)=\varepsilon^*_b=\varepsilon^b+1.$ It is therefore consistent to require that 
$$\frac{\delta^R\widetilde {\cal A}}{\delta \phi^a} \hat\otimes \frac{\delta^L\widetilde {\cal B}}{\delta \phi^*_b}=(-1)^{(\varepsilon(\widetilde {\cal A})+\varepsilon^a)(\varepsilon(\widetilde {\cal B})+\varepsilon^b+1)}\frac{\delta^L\widetilde {\cal B}}{\delta \phi^*_b} \hat\otimes \frac{\delta^R\widetilde {\cal A}}{\delta \phi^a}$$
and consequently these partial functional derivatives have the same symmetry properties as the usual ones defined for fields $\phi$ having their values in a vector space. We have successfully encoded the correct properties for vector-bundle valued fields. 
 
It follows from these remarks that we can define an anti-bracket of $\widetilde {\cal A},\widetilde {\cal B},$ $(\widetilde {\cal A},\widetilde {\cal B}):\Gamma_U\rightarrow C^{\infty}_{(\varepsilon,\varepsilon')}U^*$ by the invariant formula
$$(\widetilde {\cal A},\widetilde {\cal B})= 
\frac {\delta^R\widetilde {\cal A}}{\delta \phi^a}\hat\otimes \frac {\delta^L\widetilde {\cal B}}{\delta \phi^*_a}-
\frac {\delta^R\widetilde {\cal A}}{\delta \phi^*_a}\hat\otimes \frac {\delta^L\widetilde {\cal B}}{\delta \phi^a}.$$ 
We have imposed parities in such a manner that properties such as 
$$(\widetilde {\cal A},\widetilde {\cal B})=-(-1)^{(\varepsilon(\widetilde {\cal A})+1)(\varepsilon( \widetilde {\cal B})+1)}(\widetilde {\cal B},\widetilde {\cal A})$$
and the graded Jacobi identity
$$(-1)^{(\varepsilon(\widetilde {\cal A})+1)(\varepsilon( \widetilde {\cal C})+1)}
(\widetilde {\cal A},(\widetilde {\cal B},\widetilde {\cal C})) +$$$$
(-1)^{(\varepsilon(\widetilde {\cal B})+1)(\varepsilon( \widetilde {\cal A})+1)}
(\widetilde {\cal B},(\widetilde {\cal C},\widetilde {\cal A})) +
(-1)^{(\varepsilon(\widetilde {\cal C})+1)(\varepsilon( \widetilde {\cal B})+1)}
(\widetilde {\cal C},(\widetilde {\cal A},\widetilde {\cal B}))=0 $$
follow from computations strictly analogous to that of the usual case.

Recall that all these considerations are valid on an arbitrary open set $U\in{\cal U}.$
Thus if we now assume that $\widetilde {\cal A},\widetilde {\cal B}\in {\cal R},$ then it is appropriate to label $\widetilde {\cal A},\widetilde {\cal B}$ with a subscript $U.$
We write 
$$(\widetilde {\cal A},\widetilde {\cal B})_{\alpha}=(\widetilde {\cal A},\widetilde {\cal B})_{U_{\alpha}}=(\widetilde {\cal A}_{U_{\alpha}},\widetilde {\cal B}_{U_{\alpha}})=
(\widetilde {\cal A}_{\alpha},\widetilde {\cal B}_{\alpha})$$
for this anti-bracket on $\Gamma_{U_{\alpha}}.$ If $\widetilde {\cal A}\in {\cal R}_{\varepsilon(k)},\widetilde {\cal B}\in {\cal R}_{\varepsilon (l)},$ we have
$$\widetilde {\cal A}(\tilde\phi)(g)=\sum_{\alpha}\widetilde {\cal A}_{\alpha}(f_{\alpha}\tilde \phi)(g|U_{\alpha})\quad \quad
\widetilde {\cal B}(\tilde\phi)(h)=\sum_{\alpha}\widetilde {\cal B}_{\alpha}(f_{\alpha}\tilde \phi)(h|U_{\alpha})$$
where $\{\widetilde {\cal A}_{\alpha}\}$ and $\{\widetilde {\cal B}_{\alpha}\}$ are appropriate families of tensor-valued functionals, $\tilde\phi\in \Gamma$, and $g\in C^{\infty}M^k, h\in C^{\infty}M^l.$
As before
$$\widetilde {\cal A}_{\alpha}, \widetilde {\cal B}_{\alpha}:\Gamma_{\alpha}\rightarrow C^{\infty}_{\varepsilon(r)}U_{\alpha}^*$$
are defined on the space $\Gamma_{\alpha}$ of sections of $W|U_{\alpha}\rightarrow U_{\alpha},$ for each $\alpha$ and for $r=k,r=l,$ respectively. We now define 
$$(\widetilde {\cal A}, \widetilde {\cal B})(\tilde\phi)=\sum_{\alpha}(\widetilde {\cal A}_{\alpha}, \widetilde {\cal B}_{\alpha})(f_{\alpha}\tilde\phi)$$
for $\phi \in \Gamma_{E\oplus E^*}.$ 
Since $$(\widetilde {\cal A}, (\widetilde {\cal B}, \widetilde {\cal C}))(\phi)=\sum_{\alpha}(\widetilde {\cal A}_{\alpha}, (\widetilde {\cal B}, \widetilde {\cal C})_{\alpha})(f_{\alpha}\tilde\phi)=
\sum_{\alpha}(\widetilde {\cal A}_{\alpha}, (\widetilde {\cal B}_{\alpha}, \widetilde {\cal C}_{\alpha}))(f_{\alpha}\tilde\phi)$$
for all $\phi\in \Gamma$ we see that the graded Jacobi identity holds on ${\cal R}$ and that the bracket is graded skew-symmetric.
This concludes the proof that there is a well-defined anti-bracket on ${\cal R}.$

Now consider the Laplacian. If the tensor-valued functional $\widetilde {\cal A} \in {\cal R}_{\varepsilon(k)}$ is given by 
$$\widetilde {\cal A}(\tilde\phi)(g)=\sum_{\alpha}\widetilde {\cal A}_{\alpha}(f_{\alpha} \tilde\phi)(g|U_{\alpha})$$
as above we define
$$\Delta\widetilde {\cal A}(\tilde\phi)=
\sum_{\alpha}\Delta_{\alpha}\widetilde {\cal A}_{\alpha}(f_{\alpha}\tilde\phi)$$
where 
$$\Delta_{\alpha} \widetilde {\cal A}_{\alpha}(\tilde\phi)=
(-1)^{\varepsilon^a+1}\frac {\delta^R }{\delta \phi^a}( \frac {\delta^R \widetilde {\cal A}}{\delta \phi^*_a}) (\tilde\phi).$$
To see that the latter is independent of the choice of local sections of $E\oplus E^*\rightarrow M$
let $\{e_a\}, \{\bar e_b\}$ be two bases of sections of $E\oplus E^*)|_{U_{\alpha}}\rightarrow U_{\alpha}.$ We drop the R on $\delta^R$ in the following calculation and ignore the parity as it plays no role in the calculation. Observe that
$\frac {\delta \widetilde {\cal A}}{\delta \phi^*_a}$ is a mapping from $\Gamma_{\alpha}$ into $C^{\infty}_{\varepsilon(k)}U_{\alpha}^*$ and so $\frac {\delta }{\delta \phi^a}$ can operate on it. Also we already know how $\frac {\delta \widetilde {\cal A}}{\delta \phi^*_a}$  transforms. We have

\begin{eqnarray*}
 \frac {\delta }{\delta \bar\phi^a}( \frac {\delta \widetilde {\cal A}}{\delta \bar\phi^*_a})(\psi) &=&
D_{\psi}( \frac {\delta \widetilde {\cal A}}{\delta \bar \phi^*_a})(\bar e_a)  \\ &=&
D_{\psi}(B^a_c \frac {\delta \widetilde {\cal A}}{\delta \phi^*_a})(A_a^b e_b) \\ &=&
B^a_cA_a^bD_{\psi}(\frac {\delta \widetilde {\cal A}}{\delta \phi^*_a})( e_b) \\ &=&
\frac {\delta^R }{\delta \phi^a}( \frac {\delta \widetilde {\cal A}}{\delta \phi^*_a})(\psi) .
\end{eqnarray*} 
where $A,B$ are inverse matrices. It follows that the Laplacian is well-defined on ${\cal R}$ and the theorem follows.

\end{proof}

\section{ A New Anti-Bracket of local Functionals}
In this section we introduce a new anti-bracket of local functionals and derive an interesting
double integral which clarifies how one would compute it.
 Consider two copies of ${\bf C},$ denoted ${\bf C}_0$ and ${\bf C}_1.$ Elements of ${\bf C}_0$ are called even scalars while those of ${\bf C}_1$ are odd. If ${\cal A}:\Gamma_{E\oplus E^*}\rightarrow {\bf C}$ is any functional we write ${\cal A}:\Gamma_{E\oplus E^*}\rightarrow {\bf C}_{\varepsilon}$ if and only if ${\cal A}$ has parity $\varepsilon\in \{0,1\}.$ Given functionals
$${\cal A}:\Gamma_{E\oplus E^*}\rightarrow {\bf C}_{\varepsilon} \quad\quad
{\cal B}:\Gamma_{E\oplus E^*}\rightarrow {\bf C}_{\varepsilon'}$$ and $\tilde\phi\in \Gamma$ we define
$$({\cal A}, {\cal B})_{\tilde\phi}=(\widetilde {\cal A}_{\phi},\widetilde {\cal B}_{\phi}).$$
Thus this bracket of functionals has its values in $C^{\infty}_{\varepsilon}M^*\hat\otimes C^{\infty}_{\varepsilon'}M^*$ which is clearly different from the usual anti-bracket. On the other hand this fits into our general scheme.

Indeed the space $({\bf C}_0\oplus {\bf C}_1)\oplus \bigoplus_k
\bigoplus_{(\varepsilon_1,\cdots,\varepsilon_k)}
C^{\infty}_{(\varepsilon_1,\cdots,\varepsilon_k)}M^*$ is an algebra over {\bf C}. Our tensor- valued functionals are mappings from $\Gamma$ into this algebra and ordinary functionals are also in this space.

At this point we wish to calculate in detail this new bracket of two functionals. To do this first recall  that, for $\alpha\in C^{\infty}_{\varepsilon}M^*,\beta\in C^{\infty}_{\varepsilon'}M^*$ with $\varepsilon,\varepsilon' \in \{0,1\},$ we have that  $(\alpha\hat\otimes \beta)(f,g)=
\alpha(f)\beta(g)+(-1)^{|\varepsilon||\varepsilon'|}\alpha(g)\beta(f),$ for $f,g\in C^{\infty}M.$
We will be able to use this to obtain an explicit formula for our brackets of functionals below.\newline

If ${\cal A},{\cal B}$ are functionals we have from our remark above that at fixed $\tilde\phi\in \Gamma=\Gamma_{E\oplus E^*}$
$$({\cal A}, {\cal B})_{\tilde\phi}(\tilde\delta)=(\widetilde {\cal A}_{\tilde\phi},\widetilde {\cal B}_{\tilde\phi})(\tilde\delta)=\sum_{\alpha}(\widetilde {\cal A}_{\alpha}, \widetilde {\cal B}_{\alpha})(f_{\alpha}\tilde\delta)$$
where
$$\widetilde {\cal A}_{\tilde\phi}(\tilde\delta)(g)=\sum_{\alpha}\widetilde {\cal A}_{\alpha}(f_{\alpha} \tilde \delta)(g|U_{\alpha})\quad \quad
\widetilde {\cal B}_{\tilde\phi}(\tilde\delta)(g)=\sum_{\alpha}\widetilde {\cal B}_{\alpha}(f_{\alpha}\tilde\delta)(g|U_{\alpha})$$
and the $\tilde\phi$ are suppressed in the terms involving $\widetilde {\cal A}_{\alpha}$ and $\widetilde {\cal B}_{\alpha}.$

Moreover, since ${\cal A},{\cal B}$ are local, there exists $A,B\in Loc_{E\oplus E^*}$ such that ${\cal A}=\int_MA\nu,{\cal B}=\int_M B\nu$ and the tensor-valued functionals $\widetilde {\cal A}_{\alpha}, \widetilde {\cal B}_{\alpha}$ are given locally by

$$\widetilde {\cal A}_{\alpha}(\psi)(g)=
\int_{U_{\alpha}}[\widetilde E(A_{\alpha})\circ ( j\tilde\phi)|(U_{\alpha})](g\psi) \nu $$ and $$ \widetilde {\cal B}_{\alpha}(\psi)(g)=
\int_{U_{\alpha}}[\widetilde E(B_{\alpha})\circ ( j\tilde\phi)|(U_{\alpha})](g\psi) \nu$$
where $A_{\alpha}= A| (\pi^{\infty}_M)^{-1}(U_{\alpha}),B_{\alpha}= B| (\pi^{\infty}_M)^{-1}(U_{\alpha}),
g\in C^{\infty}U_{\alpha}, \psi\in \Gamma_{\alpha}$ and where $\Gamma_{\alpha}$ is the space of sections of $(E\oplus E^*)|U_{\alpha}\rightarrow U_{\alpha}.$ Here $\pi^{\infty}_M$ is the bundle mapping $J^{\infty}(E\oplus E^*)\rightarrow M.$

Notice that for $P\in Loc_{E\oplus E^*}$ and $1\leq a,b \leq n,$
$$\widetilde E_a(P)=(-D)_I ( \frac{\partial P}{\partial u^a_I }) =E_aP\quad  \quad \quad
\widetilde E_{b+n}(P)=(-D^*)_I  (\frac{\partial P}{\partial u^*_{bI }})=E^b_*P$$
where
$$D_{\mu}=\frac{\partial }{\partial x^{\mu}}+u^a_{\mu I}\frac{\partial }{\partial u^a_I}  \quad \quad \quad
D^*_{\mu}=\frac{\partial }{\partial x^{\mu}}+u^*_{\mu bI}\frac{\partial }{\partial u^*_{bI}} $$ and where $n$ is the dimension of a typical fiber of $E.$

To compute our  new anti-bracket of ${\cal A},{\cal B}$ we need only compute $(\widetilde {\cal A}_{\alpha},\widetilde {\cal B}_{\alpha}),$ where $(\widetilde {\cal A}_{\alpha},\widetilde {\cal B}_{\alpha})$ is the mapping from $\Gamma_{U_{\alpha}}$ into $C^{\infty}_{(\varepsilon,\varepsilon')}U_{\alpha}^*$  given  by the invariant formula
$$(\widetilde {\cal A}_{\alpha},\widetilde {\cal B}_{\alpha})= 
\frac {\delta^R\widetilde {\cal A}_{\alpha}}{\delta \tilde\phi^a}\hat\otimes \frac {\delta^L\widetilde {\cal B}_{\alpha}}{\delta \tilde\phi^*_a}-
\frac {\delta^R\widetilde {\cal A}_{\alpha}}{\delta\tilde \phi^*_a}\hat\otimes \frac {\delta^L\widetilde {\cal B}_{\alpha}}{\delta \tilde\phi^a}.$$ 
As an example, one of the four terms we need is  $$\frac {\delta^L\widetilde {\cal B}_{\alpha}}{\delta \tilde\phi^*_a}(\delta)(f)=D_{\delta}\widetilde {\cal B}_{\alpha}(e^a)(f)$$
where $f\in C^{\infty}U_{\alpha}$ and $\{e_a\}$ is a fixed basis of local sections of $\Gamma$ defined on $U_{\alpha}$ (with dual basis $\{e^b\}$).  Now $\widetilde {\cal B}_{\alpha}$  depends implicitly on $\tilde\phi,$ but is linear, so $D_{\delta}\widetilde {\cal B}_{\alpha}=\widetilde {\cal B}_{\alpha}$ is independent of $\delta.$ Thus
$$D_{\delta}\widetilde {\cal B}_{\alpha}(e^a)(f)=\int_{U_{\alpha}}(E_*B_{\alpha}\circ j\tilde \phi)(fe^a)\nu$$ and 
$$\frac {\delta^L\widetilde {\cal B}_{\alpha}}{\delta \tilde\phi^*_a}(\delta)(f)=\int_{U_{\alpha}}(E^a_*B_{\alpha}\circ j\tilde \phi)f\nu.$$

This is the correct expression for the left functional derivative but the right functional derivative will have a sign depending on parity. We define the left Euler operator $E^L$ to be the usual Euler operator  but when we need right derivatives we write $E^R=\pm E$ with the appropriate parity built in for right derivatives. 

First notice that

$$(\frac {\delta^R\widetilde {\cal A}_{\alpha}}{\delta \tilde \phi^a}\hat\otimes \frac {\delta^L\widetilde {\cal B}_{\alpha}}{\delta \tilde\phi^*_a})(\delta)(f\otimes g)=
\int_{U_{\alpha}}(E^R_aA_{\alpha}\circ j\phi)f \nu
\int_{U_{\alpha}}( (E^L)^a_*B_{\alpha}\circ j \phi^*)g \nu
+$$
$$(-1)^{|\varepsilon||\varepsilon'|}\int_{U_{\alpha}}( E^R_aA_{\alpha}\circ j\phi)g \nu
\int_{U_{\alpha}}((E^L)^a_*B_{\alpha}\circ j \phi^*)f \nu.
$$
Given two mappings $r:X\rightarrow P$ and $s:Y\rightarrow Q ,$ define a new mapping $r\times s$
from $X\times Y$ to $P\times Q$  by $(x,y)\rightarrow (r(x),s(y)).$ If $P=Q$ is the set of real or complex scalars define $r\otimes s$  by $(r\otimes s)(x,y)=r(x)s(y).$

Using this notation we {\it formally} define $$E^R_aA_{\alpha}\hat\otimes (E^L)^a_*B_{\alpha}=(E^R_aA)\otimes((E^L)^a_*B_{\alpha})+(-1)^{|\varepsilon||\varepsilon'|}((E^R_aA_{\alpha})\otimes ((E^L)^a_*B_{\alpha}))$$
so that
$$(\frac {\delta^R\widetilde {\cal A}_{\alpha}}{\delta \tilde \phi^a}\hat\otimes \frac {\delta^L\widetilde {\cal B}_{\alpha}}{\delta \tilde\phi^*_a})(\delta)(f\otimes g)=
\int \int [(E^R_aA_{\alpha}\hat\otimes (E^L)^a_*B_{\alpha})\circ (j\phi\times j\phi^*)](f\otimes g)(x,y)
( \nu_x\otimes\nu_y).$$ 
If we define a {\it formal ``bracket" } $\{\cdot,\cdot \}$ on $J^{\infty}(E\oplus E^*)$ by
$$\{A,B\}=[E^R_aA_{\alpha}\hat\otimes (E^L)^a_*B_{\alpha}]-[(E^R)_*^aA_{\alpha}\hat\otimes E^L_aB_{\alpha}],$$
then we obtain the local formula:

$$(\widetilde {\cal A}_{\alpha},\widetilde {\cal B}_{\alpha})(\delta)(f\otimes g)=
\int\int [\{A_{\alpha},B_{\alpha}\}\circ (j\phi \times j\phi^*)](f \otimes g)(x,y)(\nu_x\otimes \nu_y).$$

\noindent{\bf Remark.} While the latter formula is somewhat heuristic, in the bosonic case there is a well-defined bracket $\{\cdot,\cdot\}$ of the type referred to above although it is not a Lie bracket. In that case it is essentially the second term of an sh-Lie structure (see \cite{AF05, BFLS98, BH96} and references therein). It is likely that this is also true in this case but that aspect has not been fully explored at this time and is  beyond the scope of the present paper.

\providecommand{\bysame}{\leavevmode\hbox to3em{\hrulefill}\thinspace}

\end{document}